\documentstyle[12pt,epsf]{article}
\def\ra{\rightarrow}
\def\prd#1#2#3{{\it Phys. Rev.} {\bf D#1} #2 (19#3)}
\def\pl#1#2#3{{\it Phys. Lett.} {\bf #1B} #2 (19#3)}
\def\np#1#2#3{{\it Nucl. Phys.} {\bf B#1} #2 (19#3)}

\def\beq{\begin{equation}}
\def\eeq{\end{equation}}

\def\beqn{\begin{eqnarray}}
\def\eeqn{\end{eqnarray}}
\relax
\begin{document}
\begin{titlepage}
\def\ba{\begin{array}}
\def\ea{\end{array}}
\def\thefootnote{\fnsymbol{footnote}}
\begin{flushright}
        BINP-95-02\\
        UCD-95-33\\
        ISU-HET-95-5 \\
	October 1995
\end{flushright}
\vfill
\begin{center}
{\large \bf Study of Anomalous Couplings
at a $500$~GeV $e^+e^-$ Linear Collider
with Polarized Beams}
\vfill
	{\bf A.~A.~Likhoded$^{(a)}$,}
	{\bf T.~Han$^{(b)}$}
        {\bf  and G.~Valencia$^{(c)}$}\\
{\it  $^{(a)}$ On leave of absence from the Branch of The Institute
for  Nuclear Physics,\\
Protvino,  142284 Russia}. E-mail: likhoded@mx.ihep.su \\
{\it  $^{(b)}$ Department of Physics,
University of California at Davis, Davis~~CA 95616}\\
E-mail:  than@ucdhep.ucdavis.edu \\
{\it  $^{(c)}$ Department of Physics,
               Iowa State University,
               Ames IA 50011}\\
E-mail: valencia@iastate.edu \\
\vfill
\end{center}
\begin{abstract}

We consider the possibility of observing deviations from the
Standard Model gauge-boson self-couplings at a future
$500$~GeV $e^+ e^-$ linear collider.
We concentrate on the case in which the electroweak
symmetry breaking sector is strongly interacting and there
are no new resonances within reach of the collider. We find
a sensitivity to the anomalous couplings that is two orders
of magnitude higher than that achievable at LEP II. We also show
how a polarized electron beam extends the reach of the collider,
allowing experiments to probe different directions in parameter
space.

\end{abstract}

\end{titlepage}

\clearpage

\section{Introduction}

\noindent
The Standard Model of electroweak interactions is in remarkable agreement
with all precision measurements performed thus far \cite{reviewsm}.
These measurements, however, have not probed directly energy scales higher
than a few hundred GeV, and precise measurements have been limited to scales
up to the $Z$-mass. This has been used as a motivation to propose tests
of the Standard Model by studying the self-couplings of the electroweak
gauge bosons in future colliders.

Deviations from the self-couplings predicted by the minimal Standard Model
are called ``anomalous'' gauge boson couplings and have been studied
extensively in recent years. In particular, they have been discussed in
the context of future $e^+e^-$ colliders by many authors \cite{boud,group}.
There are two main differences between our present study and those that
can be found in the literature. We interpret the success of the Standard
Model as an indication that the $SU(2)_L\times U(1)_Y$ gauge theory of
electroweak interactions is essentially correct, and that the only sector
of the theory that has not been probed experimentally is the
electroweak symmetry  breaking sector.
This point of view has many practical consequences in
limiting the number of anomalous couplings that need to be studied, and
in estimating their possible magnitude \cite{bdv}. A second
difference with other studies, is that we consider the effect of having
polarized beams.

This paper is organized as follows.
In Section 2 we summarize the effective Lagrangian formalism that we
use to describe the anomalous couplings.
In Section 3 we apply these results to a $500$~GeV linear
collider with polarized beams and discuss the relevant phenomenology.
Finally we present our conclusions.

\section{Anomalous Couplings for a Strongly-Interacting
Electroweak Symmetry Breaking  Sector}
\noindent
We wish to describe the electroweak symmetry breaking sector in the case
in which there is no light Higgs boson or any other new particle. To do
this in a model independent manner we use an effective Lagrangian for the
interactions of gauge bosons of an $SU(2)_L \times U(1)_Y$ gauge
symmetry spontaneously broken to $U(1)_Q$.
The lowest order effective Lagrangian contains a gauge invariant mass term
as well as the kinetic terms for the $SU(2)_L$ and $U(1)_Y$ gauge bosons
\cite{longo}:
\begin{equation}
{\cal L} ^{(2)}=\frac{v^2}{4}\mbox{Tr}\biggl(D^\mu \Sigma^{\dagger} D_\mu
\Sigma
\biggr)
-\frac{1}{2}\mbox{Tr}\biggl(W^{\mu\nu}W_{\mu\nu}\biggr)
-\frac{1}{2}\mbox{Tr}\biggl(B^{\mu\nu}B_{\mu\nu}\biggr)\; .
\label{lagt}
\end{equation}
$W_{\mu\nu}$ and $B_{\mu\nu}$ are
the $SU(2)_L$ and $U(1)_Y$  field strength tensors
\begin{eqnarray}
W_{\mu\nu}&=&{1 \over 2}\biggl(\partial_\mu W_\nu -
\partial_\nu W_\mu + {i \over 2}g[W_\mu, W_\nu]\biggr)\:,
\nonumber \\
B_{\mu\nu}&=&{1\over 2}\biggl(\partial_\mu B_\nu-\partial_\nu B_\mu\biggr)
\tau_3\:,
\end{eqnarray}
and $W_\mu \equiv W^i_\mu \tau_i$. The Pauli matrices $\tau_i$
are normalized so that $Tr(\tau_i\tau_j)=2\delta_{ij}$.

The matrix $\Sigma \equiv \exp(i\vec{\omega}\cdot \vec{\tau} /v)$ contains the
would-be Goldstone bosons $\omega_i$ that give the $W$ and $Z$ their
mass via the Higgs mechanism, and the $SU(2)_L \times U(1)_Y$
covariant derivative is given by:
\begin{equation}
D_\mu \Sigma = \partial_\mu \Sigma +{i \over 2}g W_\mu^i \tau^i\Sigma
-{i \over 2}g^\prime B_\mu \Sigma \tau_3\:.
\label{covd}
\end{equation}
The physical masses
are obtained with $v \approx 246$~GeV. This non-linear realization
of the symmetry breaking sector is a non-renormalizable theory
that is interpreted as an effective field theory, valid below
some scale $\Lambda \leq 3$~ TeV. The lowest order interactions between
the gauge bosons and fermions are the same as those in the minimal
Standard Model.

Deviations from these minimal couplings (referred to as
anomalous gauge boson couplings),
correspond to higher dimension ($SU(2)_L\times U(1)_Y$ gauge invariant)
operators. For energies below the scale of symmetry breaking $\Lambda$,
it is possible to organize the effective
Lagrangian in a way that corresponds to an expansion of scattering
amplitudes in powers of $E^2/\Lambda^2$. The next to leading order
effective Lagrangian that arises in this context
has been discussed at
length in the literature
\cite{longo,holdom,fls,bdv,appel}. The
contributions of this Lagrangian to the anomalous couplings have
also been written down before \cite{appel}.

In this paper we consider the process $e^+e^- \ra W^+ W^-$ at tree
level and work in unitary gauge, therefore, the anomalous couplings
enter the calculation only through
the three gauge boson vertex $VW^+W^-$ (where $V=Z,\gamma$).\footnote{
The anomalous couplings also affect the $e\nu W$ and $e^+e^-Z$ vertices
through renormalization. However, they do so only through the parameter
$L_{10}$, and we will argue later that it is not necessary to consider
this coupling in detail because it has already been severely constrained
at LEP.}
It is conventional to write the most general  $CP$  conserving
$VW^+W^-$ vertex in the form \cite{hagi}:
\begin{eqnarray}
{\cal L}_{WWV}&= &
-ie {c_\theta \over s_\theta } g_1^Z \biggl(
W_{\mu\nu}^{\dagger} W^{\mu}-W_{\mu\nu} W^{\mu~\dagger}\biggr) Z^\nu
-ie g_1^\gamma\biggl(
W_{\mu\nu}^{\dagger} W^{\mu}-W_{\mu\nu} W^{\mu~\dagger}\biggr) A^\nu
\nonumber \\ &&
-ie {c_\theta \over s_\theta } \kappa_Z
W_{\mu}^{\dagger} W_{\nu}Z^{\mu\nu}
-ie \kappa_\gamma W_{\mu}^{\dagger} W_{\nu}A^{\mu\nu}
\nonumber \\ & &
-e {c_\theta  \over s_\theta } g_5^Z
\epsilon^{\alpha\beta\mu\nu}\biggl(
W_\nu^-\partial_\alpha W_\beta^+-W_\beta^+\partial_\alpha
W_\nu^-\biggr)Z_\mu \;. \label{vertx}
\end{eqnarray}
where $s_\theta=\sin \theta^{}_W, c_\theta =\cos \theta^{}_W$.
The effective Lagrangian framework for the case of a strongly interacting
symmetry breaking sector, predicts the five constants in Eq.~(\ref{vertx}),
they are \cite{appel,valen2}:
\begin{eqnarray}
g_1^Z&=&1+{e^2\over c_\theta^2}
\biggl({1\over 2 s_\theta^2 } L_{9L}
+{1\over  (c_\theta^2-s_\theta^2)}L_{10}\biggr){v^2\over
\Lambda^2}+\cdots \;,\nonumber \\
g_1^\gamma&=& 1+\cdots \;, \nonumber \\
\kappa_Z&=&1+ e^2\biggl({1\over 2 s_\theta^2
c_\theta^2} \biggl(L_{9L}c_\theta^2
-L_{9R}s_\theta^2\biggr)
+{2 \over  (c_\theta^2-s_\theta^2)}L_{10}
\biggr){v^2\over \Lambda^2}+\cdots \;,  \label{unot}    \\
\kappa_\gamma&=&1+{e^2 \over s_\theta^2}
\biggl({L_{9L}+L_{9R}\over 2} -L_{10}\biggr)
{v^2\over \Lambda^2}+ \cdots\;, \nonumber \\
g_5^Z &=& {e^2 \over s_\theta^2
c_\theta^2}\hat{\alpha}{v^2\over \Lambda^2}+\cdots\;. \nonumber
\end{eqnarray}
In Eq.~(\ref{unot}) we have written down the leading contribution
to each anomalous coupling,\footnote{This is why we do not have terms
corresponding to the usual $\lambda_Z$ and $\lambda_\gamma$: they
only occur at higher order in $1/\Lambda^2$.}
and denoted by $\cdots$ other contributions
that arise at higher order (${\cal O}(1/\Lambda^4)$), or at order
${\cal O}(1/\Lambda^2)$ with custodial $SU(2)$ breaking. We are
thus assuming that whatever breaks electroweak symmetry has at
least an approximate custodial symmetry. Under these assumptions
there are only four operators in the next to leading order effective
Lagrangian that are relevant:
\begin{eqnarray}
{\cal L} ^{(4)}\ &=&\ \frac{v^2}{\Lambda^2}  \biggl\{
 - i g L_{9L} \,\mbox{Tr}\biggl( W^{\mu \nu} D_\mu
\Sigma D_\nu \Sigma^{\dagger}\biggr)
\ -\ i g^{\prime} L_{9R} \,\mbox{Tr} \biggl(B^{\mu \nu}
D_\mu \Sigma^{\dagger} D_\nu\Sigma\biggr) \nonumber \\
& +& g g^{\prime} L_{10}\, \mbox{Tr}\biggl( \Sigma
B^{\mu \nu}
\Sigma^{\dagger} W_{\mu \nu}\biggr)\ +\
g {\hat \alpha}
\epsilon^{\alpha \beta
\mu \nu}\mbox{Tr}\biggl(\tau_3 \Sigma^{\dagger} D_\mu \Sigma\biggr)
\mbox{Tr}\biggl( W_{\alpha \beta} D_\nu \Sigma \Sigma^{\dagger}\biggr)
\biggr\} \label{lfour}
\end{eqnarray}
The first three terms conserve the custodial
$SU(2)_C$ symmetry, and
we have explicitly introduced the factor $v^2/\Lambda^2$ in our definition
of ${\cal L}^{(4)}$ so that the
$L_i$ are naturally of ${\cal O}(1)$. The term with $\hat{\alpha}$
breaks the custodial symmetry but we include it because it provides the
leading contribution to $g_5^Z$. In theories with a custodial symmetry,
this term is, therefore, expected to be smaller than the other ones in
Eq.~(\ref{lfour}). This term is also special in that it is the only one
at ${\cal O}(1/\Lambda^2)$ that violates parity while conserving $CP$.
With our normalization, we expect
$\hat{\alpha}$ to be of ${\cal O}(1)$ in theories without a custodial
symmetry and much smaller in theories that have a custodial
symmetry \cite{dv}.

For our discussion we will assume that the new
physics is such that the tree-level coefficients of ${\cal L}^{(4)}$ are
larger than the (formally of the same order) effects induced by
${\cal L}^{(2)}$ at one-loop. More precisely, that after using dimensional
regularization and a renormalization scheme similar to the one used
in Ref.~\cite{bdv}, the $L_i(\mu)$ evaluated at a typical scale (around
500 GeV for this process)
are equal to the tree-level coefficients, and that their scale dependence
is unimportant for the energies of interest. The physical motivation
for this assumption is that, even if we do not see any new resonances
directly, the effects of the new physics from high mass scales
must clearly stand out if there
is to be any hope of observing them.
When the indirect effects of the
new physics enter at the level of SM radiative corrections, very precise
experiments (as the ones being performed at LEP I) are needed to unravel them.
We are assuming that there will not be any such precision measurements
in the next generation of high energy colliders.

All the necessary Feynman rules in unitary gauge have been written
down in Ref.~\cite{valen2}.
For our numerical study we will use the input parameters:
\begin{equation}
M_Z = 91.187 \mbox{ GeV,\ \ }
\alpha = 1/128.8\:,\;\; G_F=1.166\cdot 10^{-5}\mbox{~GeV}^{-2}\:. \label{param}
\end{equation}
We will also use $\Lambda=2$~TeV as the scale normalizing our next to leading
order effective Lagrangian, Eq.~(\ref{lfour}).

The parameter $L_{10}$ can be very tightly constrained by
precision measurements at LEP I \cite{valen1}:
\begin{equation}
-1.1 \leq L_{10}(M_Z) \leq 1.5\:. \label{altb}
\end{equation}
We find that this bound cannot be significantly improved with a 500 GeV
linear collider so we will not study $L_{10}$ further in this paper.

To summarize, we consider the next to leading order effective Lagrangian
for a  $CP$  conserving,
strongly interacting, electroweak symmetry breaking sector with
an (at least) approximate custodial symmetry. We then find that the leading
contribution to the anomalous couplings relevant for $e^+e^- \ra W^+W^-$
at $\sqrt{s}=500$~GeV can be written down in terms of four coupling
constants. Finally we note that one of those coupling constants has
already been tightly constrained at LEP~I. We are thus left with a model
that contains only three parameters $L_{9L}$, $L_{9R}$, and $\hat{\alpha}$.
In the following sections we discuss the phenomenology of these three
constants at a future linear collider with polarized beams.

\section{Bounds from the process $e^+e^-\to W^+W^-$}

\begin{figure}[htb]
\centerline{\epsfxsize=4.5in\epsfbox{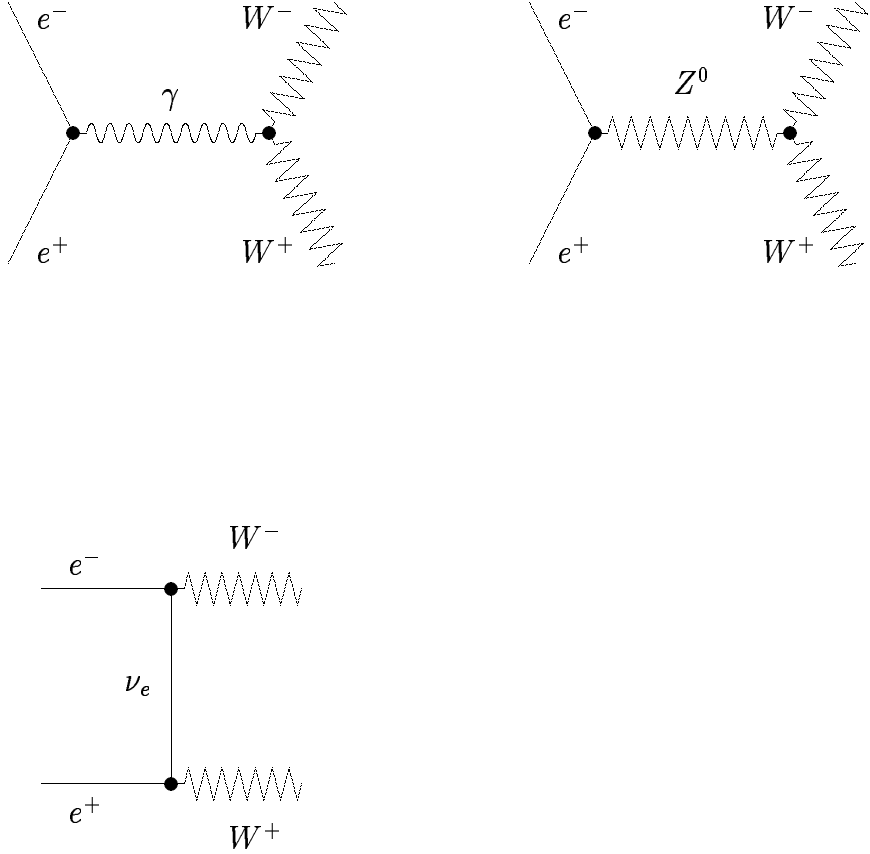}}
\caption{Diagrams contributing to the process $e^+e^-\to W^+W^-$. The full
circles represent vertices that include both the lowest order interaction
and the anomalous couplings discussed in the text.}
\label{ffr}
\end{figure}

\noindent
The process of $W$-boson pair production in $e^+ e^-$ collisions in the Born
approximation is determined by the diagrams shown in Fig.~\ref{ffr}.
The full circles
represent vertices that include both the standard model couplings, and the
anomalous couplings. The anomalous couplings enter these vertices directly
or through renormalization of standard model parameters \cite{valen2}.
We will denote the degree of longitudinal
polarization of the electron and positron by $z_1$ and $z_2$, respectively.
Our notation is such that $z_1=1$ corresponds to a {\it right-handed}
electron, whereas $z_2=1$ corresponds to a {\it left-handed} positron.
The cross section for $e^+e^-\to W^+W^-$ with polarized beams
can be written in terms of the usual Mandelstam variables $s$ and $t$ as:
\begin{eqnarray}
\int_{t_{min}}^{t_{max}}\frac{d\sigma}{dt}dt &=&
\frac{\pi\alpha^2}{4 s^2 M_W^4} \nonumber \\
&\cdot& \sum^{3}_{i,j=1}
C_{ij}\left(T_{ij}(t_{max})- T_{ij}(t_{min})\right)\:. \label{sigmat}
\end{eqnarray}
The terms $C_{ij}T_{ij}$ give the contributions of the pair products of
amplitudes of the corresponding diagrams (see Fig.~\ref{ffr})
to the cross-section.
The coefficients $C_{ij}$ depend on the electroweak
parameters and on the polarization of the initial particles. They are:
\begin{eqnarray}
C_{11} & = & \frac{S_1}{s^2}, \nonumber \\
C_{12} & = & -2\frac{(s-M_Z^2)c_\theta }{s_\theta
s ((s-M_Z^2)^2 + M_Z^2 \Gamma_Z^2)}, \nonumber \\
C_{22} & = & \frac{c^2_\theta}{s^2_\theta
 ((s-M_Z^2)^2 + M_Z^2 \Gamma_Z^2)},  \\
C_{13} & = & \frac{S_2 - S_1}{2s s^2_\theta}, \nonumber \\
C_{23} & = & \frac{(s-M_Z^2)c_\theta }{2 s^3_\theta
 ((s-M_Z^2)^2 + M_Z^2 \Gamma_Z^2)}, \nonumber \\
C_{33} & = & \frac{S_1 - S_2}{8 s^4_\theta}, \nonumber
\end{eqnarray}
where $S_1$ and $S_2$ carry the dependence on the beam polarization:
\begin{equation}
S_1 = 1+z_1 z_2,\;\;\; S_2 = z_1 + z_2\:. \label{polar}
\end{equation}

\noindent
Analytic expressions for $T_{ij}=T_{ij}(M_W,\kappa_{\gamma,Z},
g_{1\gamma,1Z},g_5,
s,t)$  are given in the Appendix. With $\theta$ the angle between
the incoming electron and the outgoing $W^-$ in the $e^+e^-$ center
of mass frame, we can use Eq.~(\ref{sigmat}) to construct the differential
cross-section and the $\cos\theta$ distribution for any angular binning.

\subsection{Assumed experimental parameters}

\noindent
In order to study the physics of anomalous couplings at a $500$~GeV
linear collider, we first need to know some machine and detector
parameters.

For the collider we will use an integrated luminosity of
$\int {\cal L}dt=50 \; fb^{-1}$ per year and a center of mass energy of
$\sqrt{s} = 500$ GeV,
the numbers commonly used for NLC, CLIC, VLEPP and JLC projects.
For the maximal degree of beam polarization we use the
values determined by the VLEPP study group \cite{vlepppol}:
$z_1,\: z_2=(-0.8,\: 0.8)$. Depending on the mechanism used
to polarize the beams it should at least be possible to achieve this
high a polarization for the electrons \cite{nopos}. This is very
encouraging because we will find
that to place bounds on the anomalous gauge boson couplings
of our model there is no need for positron polarization.

We will use the conservative estimates of Ref.~\cite{djoud,miller}
for the expected systematic errors in the measurements of  the
muonic and  hadronic cross-sections and  asymmetries, and
in the luminosity in the experiments at the  500-GeV collider:
\begin{center}
\begin{tabular}{|c|c|c|c|c|c|}\hline
  & $\Delta\epsilon_{\mu}/\epsilon_{\mu}$ &
$\Delta\epsilon_{h}/\epsilon_{h}$ & $\Delta A^l_{FB}$ & $\Delta A_{LR}$ &
$\Delta L/L$ \\ \hline
$\Delta_{syst}$ & 0.5\% & 1.\% & $\ll$1.\% & 0.003 & 1.\% \\ \hline
\end{tabular}
\end{center}
A detailed investigation of the process $e^+e^-\to W^+W^-$
has shown that the systematic error in the  cross-section measurement
can be $\sim$2\% \cite{frank,gouna,choi}.
This error is due to
the uncertainty in the luminosity measurement ($\delta {\cal L}\simeq$1\%),
the error in the acceptance ($\delta_{accep.}\simeq$1\%),
the error for background subtraction ($\delta_{backgr.}\simeq$0.5\%)
and a systematic error for the knowledge of the branching ratio
($\delta_{Br}\simeq$0.5\%).
In order to fully reconstruct the $WW$-pair events and to identify the
$W$ charges, we consider only the ``semileptonic'' channel, namely,
$WW \rightarrow  l^\pm \nu + 2$-jets.
According to the preliminary estimates
of Ref.~\cite{frank,gouna}, the efficiency for $WW$-pair
reconstruction (using the ``semileptonic'' channel) is  $\epsilon_{WW}=0.15$.
It is easy to estimate that for the anticipated luminosity of
$\sim 50\; fb^{-1}$ the expected number of unreconstructed events
is $\sim 3.7\times 10^5$, which
corresponds to a relative statistical error in the  cross-section value of
$\sim 0.17$\%. After reconstruction, the  number of $WW$-pairs is
about $\sim 5.5\times 10^4$, which
corresponds to a relative statistical error of $\sim 0.4$\%.

This means that for this process the systematic error may be the dominant
one. However, this situation could change when there are kinematical
cuts, or when the beams are polarized. To be conservative, we
thus include both the statistical error and an estimate of a
possible systematic error in our analysis.

\subsection{Observables used to bound new physics}

\noindent
The choice of experimental observables and
data processing procedure is  crucial in analyzing the
capability of the future $e^+e^-$ collider to place bounds
on new physics.
The total and differential cross-sections, as well as the asymmetries
of the process under study, are commonly used.
To discuss the sensitivity
of the $e^+e^-\to W^+W^-$ process to $L_{9L}$, $L_{9R}$, and $\hat{\alpha}$,
we will use the total cross-section
$\sigma_{total}$ and the asymmetry $A_{FB}$.
For this process these quantities are defined analogously to
the case of $e^+e^-\to f\bar f$.\footnote{Recall that we only use the
channel that allows a complete reconstruction of the $WW$ pair.}

Typically one uses the SM predictions
as the ``experimental'' data,\footnote{There are several ways for
such data modelling: a) application of the analytical SM expressions
to represent ``experimental'' distributions,
see, for example, \cite{choi}; b) Monte-Carlo simulation  of the experimental
distributions according to the SM predictions taking into account a
probabilistic spread, see, for example \cite{miy,bark}.}  and considers
possible effects due to new  physics as small
deviations. One then requires agreement between the
predictions including new physics and the ``experimental'' values within
expected experimental errors.
The parameters representing new physics are, thus, bound by requiring
that their effect on the selected observables be smaller than the
expected experimental errors.

It is common to consider differential distributions such as
$d\sigma/d\cos\theta$ as observables
(where $\theta$ is  the angle between the $e^-$-beam direction
and the direction of the $W^-$). However, as it has been emphasized in
Ref.~\cite{my}, it is difficult to perform a meaningful analysis of these
distributions in the absence of real experimental data and detailed
knowledge of the detector.
We start our analysis using the total cross-section
and forward-backward asymmetry as observables.
These two observables are constructed
from the independent measurements of the forward and backward cross-sections
$\sigma_F$ and $\sigma_B$. The two observables:
$\sigma = \sigma_F + \sigma_B$ and $\sigma\cdot A_{FB} = \sigma_F - \sigma_B$
are thus independent and we can analyze them simultaneously by requiring
that:
\begin{equation}
\sqrt{\left(\frac{\sigma -\tilde\sigma}{\Delta\sigma} \right)^2+
      \left(\frac{A_{FB} -\tilde A_{FB}}{\Delta A_{FB}} \right)^2}\leq
\;\mbox{number of standard deviations}\: . \label{total2}
\end{equation}
In this way
we use all the information in the  total cross-section,
as well as  partial information from angular dependence. In Eq.~(\ref{total2})
$\sigma\equiv \sigma^{SM}$ and $A_{FB}\equiv A_{FB}^{SM}$
represent anticipated experimental data, $\tilde\sigma$ and
$\tilde A_{FB}$ are the predictions including new physics.
$\Delta\sigma$ and $\Delta A_{FB}$ are the corresponding absolute
uncertainties including  systematic and statistical
errors.\footnote{It should be noted that for the case of $A_{FB}$ the bulk
of the systematics (for example the uncertainty due to luminosity
measurements), cancels out.}
We have:\footnote{We neglect any
correlation between statistical and systematic errors.}
\begin{eqnarray}
\Delta\sigma &=& \sigma^{SM} \cdot \sqrt{\delta_{stat}^2+\delta_{syst}^2}\:,
\label{uncer1} \\
\delta_{stat} &=& \frac{1}{\sqrt{N_{events}}}=
\frac{1}{\sqrt{\epsilon_{WW}{\cal L}\sigma^{SM}}}\:, \nonumber\\
\delta_{syst} &=& \sqrt{\delta {\cal L}^2+\delta_{accep}^2+\delta_{backgr}^2+
\delta_{Br}^2}\:, \nonumber
\end{eqnarray}
and
\begin{eqnarray}
\Delta A_{FB} &=& A_{FB}^{SM} \cdot \sqrt{\delta_{1\: stat}^2+\delta_{1\:
syst}^2}\:,
\label{uncer2} \\
\delta_{1\: stat} &=& \frac{1}{\sqrt{N_{events}}}
\sqrt{\frac{1-A_{FB}^2}{A_{FB}^2}}\:, \nonumber\\
\delta_{1\: syst} &=& \sqrt{\delta_{accep}^2+\delta_{backgr}^2+
\delta_{Br}^2}\:, \nonumber
\end{eqnarray}
A typical choice for the number of standard deviations
in Eq.~(\ref{total2}) is two.
Assuming a Gaussian distribution for the systematic errors,
this $2\sigma$ level corresponds to 95\%  C.L. for
the resulting bounds on the parameters under study.

It is possible to use more information from the angular
distribution than that present in the forward-backward asymmetry.
To do so, one can use a simple $\chi^2$-criterion defined as
\begin{equation}
\chi^2 = \sum_i \left(\frac{X_i - Y_i}{\Delta^i_{exp}}
\right)^2 , \label{chi2}
\end{equation}
where
\begin{displaymath}
X_i = \int_{\cos\theta_i}^{\cos\theta_{i+1}}
\frac{d\sigma^{SM}}{d\cos\theta}d\cos\theta,\;\;\; Y_i =
\int_{\cos\theta_i}^{\cos\theta_{i+1}}
\frac{d\sigma^{NEW}}{d\cos\theta}d\cos\theta \:,
\end{displaymath}
and $\Delta^i_{exp}$ are the corresponding (expected)
experimental errors in each
bin defined as in Eq.~(\ref{uncer1}).
For the binning we subdivide the chosen range of $\cos\theta$
into equal bins. This procedure gives us a rough idea
of the additional information present in the angular distribution. However, a
significant analysis of the angular distribution cannot really be done at
this stage as discussed in Ref.~\cite{my}.

\subsection{Bounding $L_{9L}$, $L_{9R}$ and $\hat{\alpha}$}
\noindent
In a scenario for electroweak symmetry breaking like the one discussed
in Section~2, we have only three parameters determining the
anomalous couplings: $L_{9L}$, $L_{9R}$, and $\hat{\alpha}$. This scenario is
analyzed in terms of an effective Lagrangian with operators of higher
dimension being suppressed by additional powers of the scale of new
physics $\Lambda$. Our amplitudes involving the couplings $L_{9L}$, $L_{9R}$
and $\hat\alpha$ are, thus, the lowest order terms in a perturbative
expansion in powers of $(E^2,v^2)/\Lambda^2$.
For the whole formalism to make sense, the corrections to the standard
model amplitudes (linear in the anomalous couplings) must be small.
For a numerical analysis one can take two different points of view:
\begin{itemize}
\item Formally, we have truncated the amplitudes at order $1/\Lambda^2$.
Therefore, when calculating the cross-section we must drop the terms
quadratic in the anomalous couplings since our calculation is only
complete to order $1/\Lambda^2$. We will call this approach the ``linear''
approximation.

\item We may invoke a naturalness assumption, under which we do not
expect contributions to an observable that come from different
anomalous couplings to cancel each other out. Under this assumption
we truncate the amplitudes at order $1/\Lambda^2$, but after this
we treat them as exact. We will refer to this approach as the
``quadratic'' approximation from now on.
\end{itemize}
Clearly, if the perturbative expansion is adequate, both approaches
will lead to the same conclusions; the difference between them being
higher order in the $1/\Lambda^2$ expansion. We will mostly use the
``linear'' approximation, but we will occasionally use
the ``quadratic'' approximation for comparison as well.
Any difference between them may
be considered a rough estimate of the theoretical uncertainty.

We will consider three cases: one in which the beams are unpolarized;  one
in which both electron and positron beams have their maximum degree of
polarization, $|z_{e^+,e^-}|=0.8$; and one in which only the electron beam
is polarized, $|z_{e^-}|=0.8$, $z_{e^+}=0.$

\subsubsection{Dependence on angular cut}
\noindent
The process $e^+e^-\to W^+W^-$ proceeds via the three diagrams in
Figure~\ref{ffr}.
Of these, the $t$-channel neutrino exchange diagram dominates the
cross-section. This dominant contribution to the cross-section, however,
does not depend on the new physics parameters $L_{9L}$, $L_{9R}$,
or $\hat{\alpha}$. Since this dominant contribution is peaked at small
values of the angle $\theta$, we expect to improve the sensitivity
to new physics by excluding this kinematic region.
To implement this idea we impose the cut $|\cos\theta| \leq c < 1$
and study the resulting interplay between a better sensitivity to the
anomalous couplings and a loss in the number of events
(with the corresponding increase in statistical error). We have
studied the dependence of the bounds on the kinematical cut $|\cos\theta|
\leq c$ for the range $0.1\leq c \leq 0.989$ (the upper limit
corresponding to the minimal characteristic  scattering angle
defined by the geometry of the experimental setup \cite{frank,gouna}).
We find that this symmetric kinematical cut does not affect the
bounds significantly.

\begin{figure}[htb]
\centerline{\epsfxsize=2.in\epsfbox{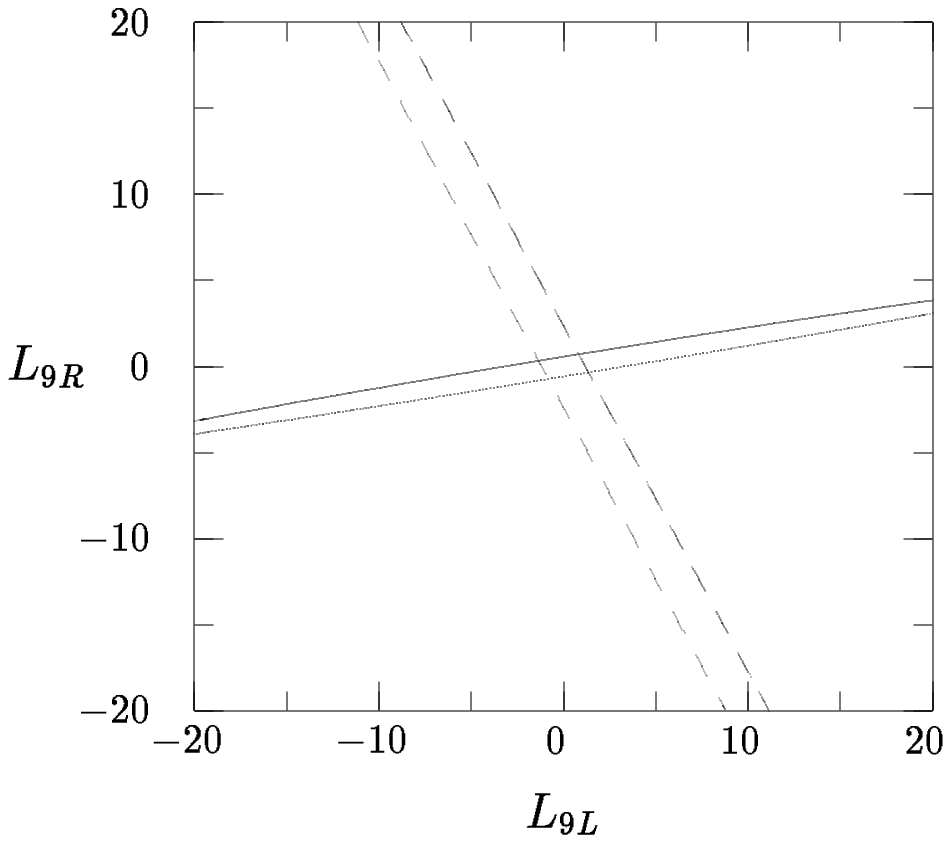}\hspace{0.5in}
\epsfxsize=2.in\epsfbox{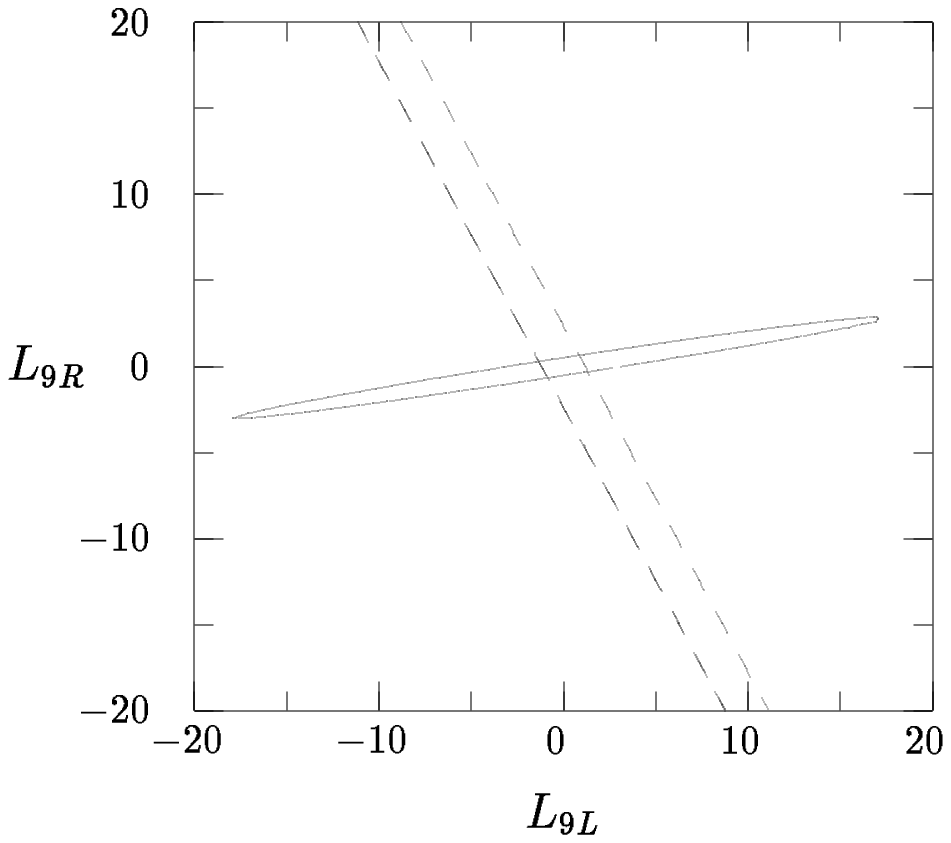}}
\centerline{\epsfxsize=2.in\epsfbox{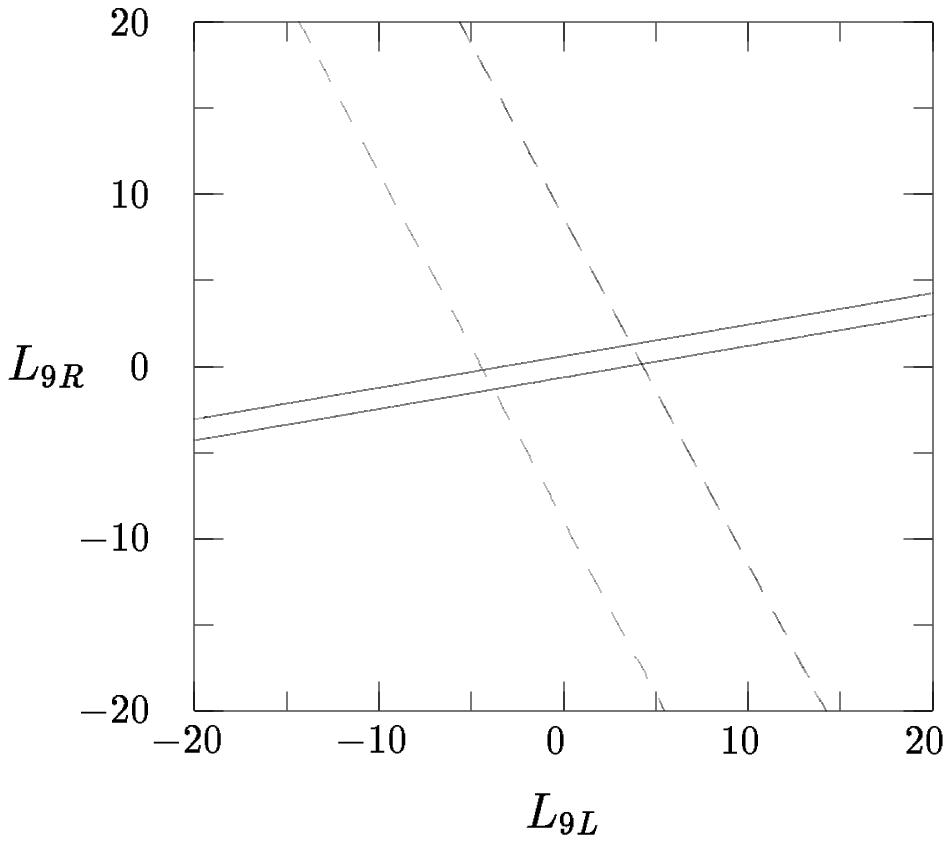}}
\caption{Allowed region for the $L_{9L}-L_{9R}$ parameters at $\hat
\alpha =0$ for the initial beam polarizations $z_1=z_2=0.8$
(solid contour) and $z_1=z_2=0$ (dashed contour)
for  cuts on the scattering angle $-0.989\leq \cos\theta
\leq c_2$, where: a) $c_2=0.1$; b) $c_2=0.4$; c)$c_2=0.989$. We use the
``linear'' approximation discussed in the text.}
\label{fac}
\end{figure}

Nevertheless, it is possible to improve the sensitivity of this
process to the anomalous couplings by using an {\it asymmetric}
kinematical cut of the form $-1\leq c_1\leq \cos\theta \leq c_2\leq 1$.
With a strong cut in the forward direction and a weak cut in the
backward hemisphere one can reduce the $t$-channel background
with a tolerable loss of statistics.
We have explored the sensitivity of the resulting bounds to the value
of the cuts for a wide range of parameters $c_1$ and $c_2$, and  for different
combinations of initial particle polarizations.
As a typical example we present in Fig.~\ref{fac}
the allowed $L_{9L}-L_{9R}$ parameter region for
unpolarized (dashed line) and maximally polarized (solid line) beams.
We set $\hat \alpha =0$, and show three sets of angular cuts for the forward
hemisphere: $c_2= 0.1,\: 0.4,\: 0.989$, while keeping $c_1=-0.989$.
We find an optimal set of cuts that we will use for
the remainder of our analysis given by:
\begin{equation}
c_1= -0.989,\: c_2 \simeq 0.4.
\label{angcuts}
\end{equation}

\subsubsection{Polarization dependence}
\noindent
An interesting question is whether the use of polarized beams
significantly improves the bounds that can be placed on the anomalous
couplings. A preliminary study in Ref.~\cite{dv} indicated that
the sensitivity to $\hat{\alpha}$ is greatly increased with
polarized beams, but only if the degree of polarization is very close to one.
Here we study the effect of having a degree of polarization that
can be achieved in practice, $z \leq 0.8$.

In Fig.~\ref{fac}b we present the allowed $L_{9L}-L_{9R}$
parameter region (with $\hat \alpha =0$) for maximally ($z_1=z_2=0.8$)
polarized and unpolarized beams.
We see that the bounds that can be obtained with polarized beams
(solid lines) are slightly better than the bounds that can be obtained
with unpolarized beams (dashed lines).
This effect is due to the reduction of the relative contribution of
the ``background'' $t$-channel diagram which results in a better sensitivity
of the process to the anomalous couplings. With the maximum degree
of polarization that can be achieved in practice, one does not find
the spectacular effects that could be found with completely polarized
beams \cite{dv}.

\begin{figure}[htb]
\centerline{\epsfxsize=2.in\epsfbox{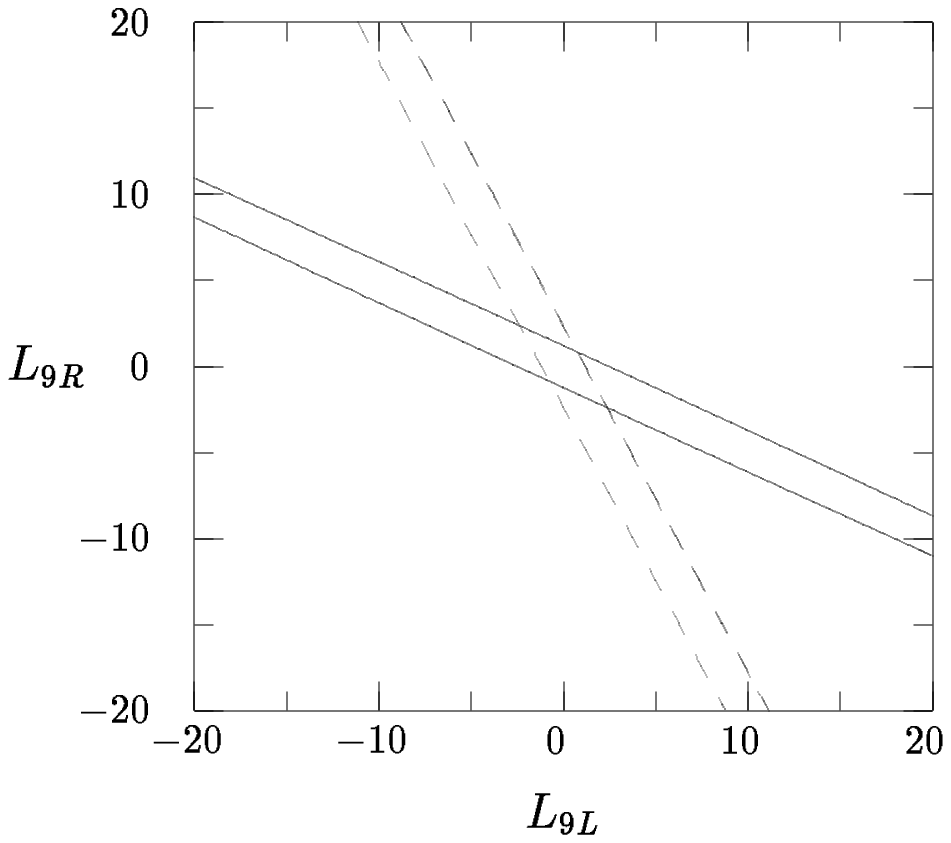}\hspace{0.5in}
\epsfxsize=2.in\epsfbox{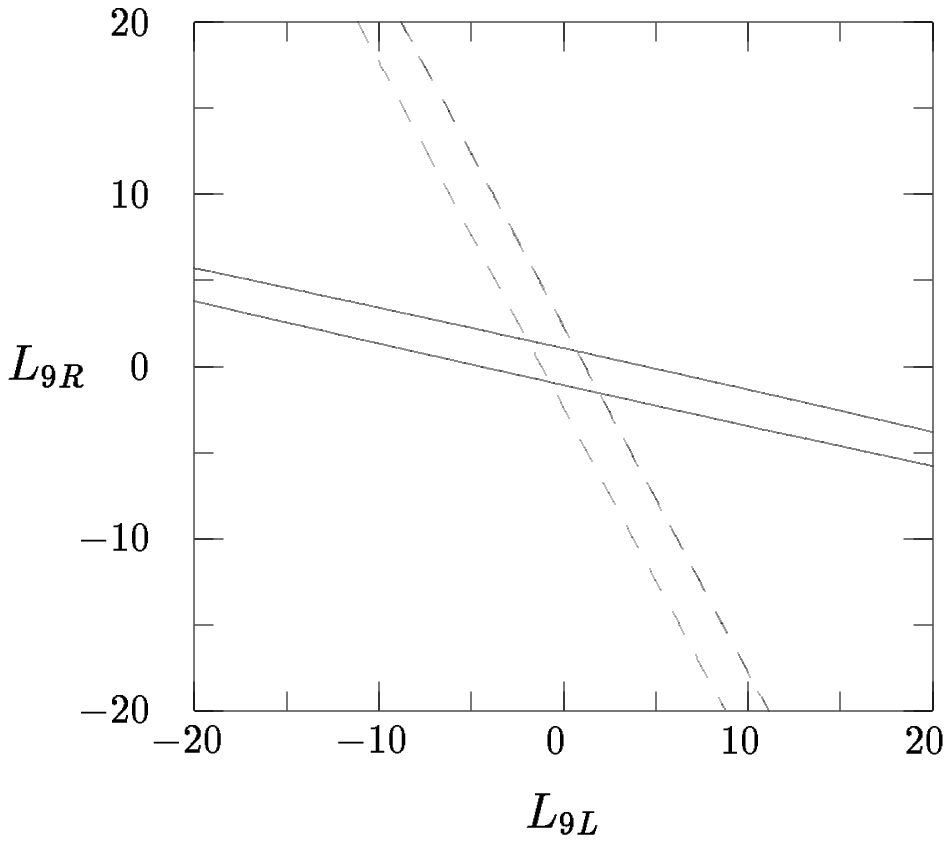}}
\caption{Allowed region for the $L_{9L}-L_{9R}$ parameters at $\hat
\alpha =0$ for cuts on the scattering angle $-0.989\leq \cos\theta
\leq 0.4$ for beam polarizations (dashed contour represents
the unpolarized case):  a) $z_1=z_2=0.4$;
b) $z_1=0.8$, $z_2=0$.
We use the ``linear'' approximation discussed in the text.}
\label{fpol}
\end{figure}

Nevertheless, polarized beams are very useful to constrain new physics
that is described by several unknown parameters.
The unpolarized case can only constrain a particular linear
combination of parameters (in this case $L_{9L}$ and $L_{9R}$) thus
giving the dashed band shown in Fig.~\ref{fac}b.
The polarized result depends on a
{\it different} linear combination of parameters. The simultaneous study
of polarized and unpolarized collisions can, therefore, give much better
bounds on the anomalous couplings than either one of them separately.

An intermediate degree of polarization, such as $z_1=z_2=0.4$ also leads
to an improvement of the bounds (see Fig.~\ref{fpol}a),
although it is not as effective as
the case with maximum practical degree of polarization in reducing the
allowed region of parameter space when combined with the unpolarized
measurement. If polarization is
available only for the electron beam it is still possible to reduce the
region of parameter space that is allowed by the unpolarized
measurement. We illustrate this in Fig.~\ref{fpol}b where we show the
case $z_1=0.8$, $z_2=0$.

Using the ``quadratic'' approximation, one finds that each allowed region
of parameter space in Fig.~\ref{fpol} is replaced by several possible regions.
This is because the terms that are quadratic in the anomalous couplings in
the cross-section give rise to allowed regions shaped like ellipsoids. The
case with polarized beams gives rise to a rotated ellipsoid, and the two
intersect in more than one region. It is obvious, however, that only the
region that contains the standard model point is physical, and this region
is very much like that shown in Fig.~\ref{fpol} for the ``linear''
approximation. It is interesting to notice that one could decide which is
the true allowed region experimentally. By changing the degree of polarization
one obtains a different rotated ellipsoid that intersects the unpolarized
one in several regions. Only the region containing the standard
model point is common to the different degrees of beam polarization.
This further illustrates the complementarity of polarized and
unpolarized measurements.

\section{Results}

\noindent
We first present the bounds on the anomalous couplings that follow from
Eq.~(\ref{total2}).
In the case of the ``quadratic'' approximation, the cross-section contains
terms that are quadratic in the anomalous couplings, as well as interference
terms between the different anomalous couplings.
The allowed parameter region is a volume element in the
$L_{9L}-L_{9R}-\hat \alpha$ space enclosed by a nontrivial surface.
Due to the interplay between couplings, the allowed volume may have
holes, and therefore, it is in general not adequate to study two
dimensional projections. In keeping with our previous discussion we
select the allowed region that contains the standard model point, and
that is very similar in shape to the results of the ``linear'' approximation.
Doing this we have a simple region for which two-dimensional projections
are adequate.

\begin{figure}[htb]
\centerline{\epsfxsize=2.in\epsfbox{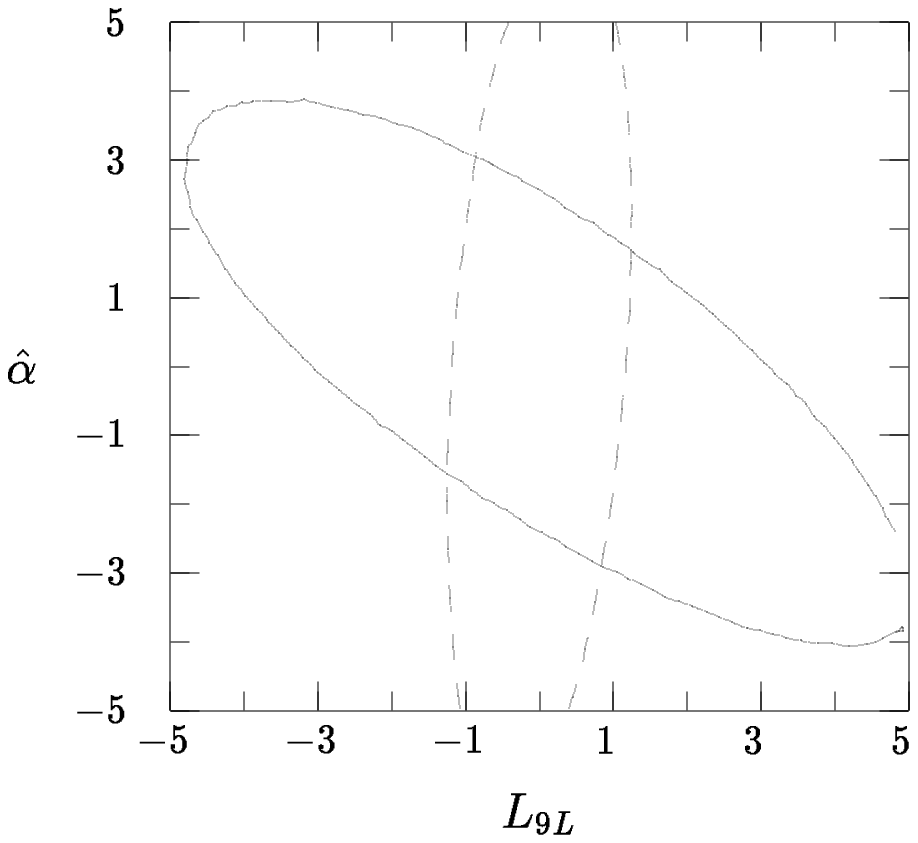}\hspace{0.5in}
\epsfxsize=2.in\epsfbox{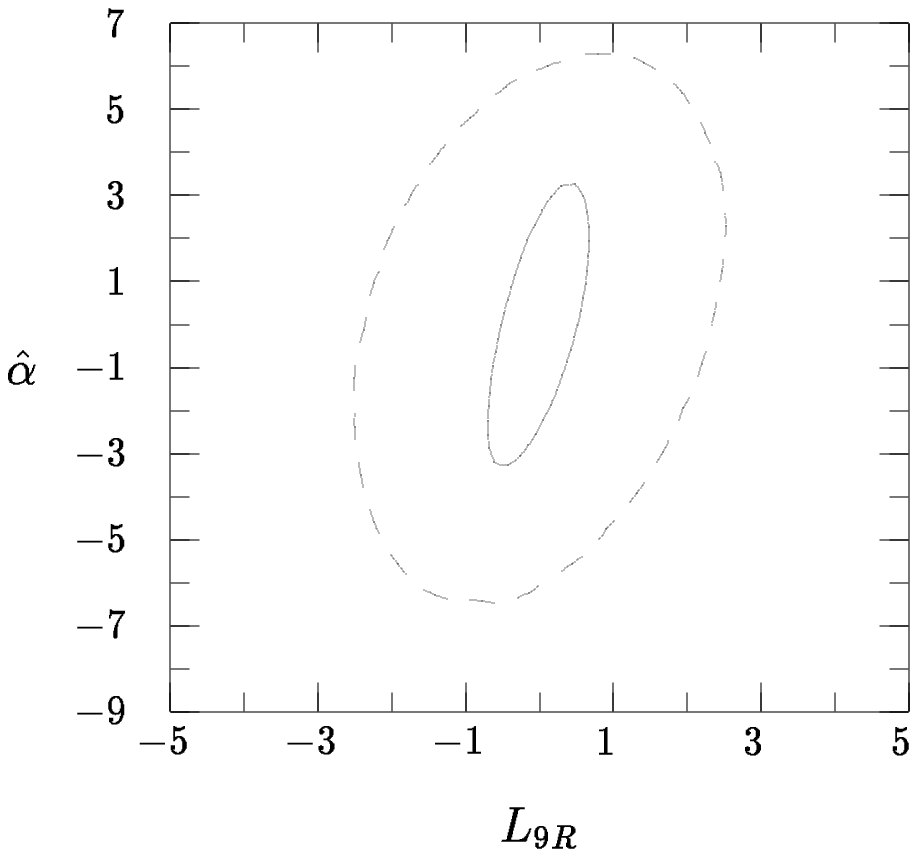}}
\caption{
Allowed regions (the case of the linear approximation)
for:
   a)  $L_{9L}-\hat \alpha$, when $L_{9R}=0$;
   b)  $L_{9R}-\hat \alpha$, when $L_{9L}=0$.
The solid contours correspond to the maximum beam polarization
$z_1=z_2=0.8$ and the dashed contours correspond to unpolarized
beams.}
\label{fpairs}
\end{figure}

We present in Fig.~\ref{fpairs} the two-dimensional projections obtained
in the directions in which one of the three anomalous couplings vanishes.
We present the case corresponding to two
standard deviation ( 95\% C.L.) bounds
from Eq.~(\ref{total2}). These results correspond to the ``linear''
approximation, but are practically identical to those obtained in
the ``quadratic'' approximation. Thus, the bounds correspond to anomalous
couplings that are small enough for the perturbative expansion to be
meaningful. This, in itself, indicates that
a 500~GeV linear collider with polarized beams will be able to place
significant bounds on a strongly interacting symmetry breaking sector.
Allowing two of the couplings to vary and setting the third one to its
standard model value we find (``linear'' case):
\begin{eqnarray}
-1.4\: \leq &L_{9L}& \leq \: 1.4\;, \nonumber\\
-0.7\: \leq &L_{9R}& \leq \: 0.7 \;, \label{bounds1} \\
-3.3 \: \leq &\hat \alpha & \leq \: 3.3\:. \nonumber
\end{eqnarray}
or (``quadratic'' case):
\begin{eqnarray}
-1.3\: \leq &L_{9L}& \leq \: 1.3\;, \nonumber\\
-0.6\: \leq &L_{9R}& \leq \: 0.7 \;, \label{bounds2} \\
-3.4 \: \leq &\hat \alpha & \leq \: 3.2\:. \nonumber
\end{eqnarray}

It is worth mentioning that the allowed regions are sometimes bound by
curved lines, even in the ``linear'' approximation. This is due to the
intrinsically non-linear combination of observables that we used,
Eq.~(\ref{total2}). In this
respect, one interesting feature can be seen in Fig.~\ref{fpairs}.
While the allowed regions in Fig.~\ref{fpairs}a and
Fig.~\ref{fpairs}b are bounded by curves, the domain
in Fig.~\ref{fac}b is bound by almost straight lines. This means that
the deviations of the $L_{9L},\: L_{9R}$ parameters affect mainly the
cross-section, but practically do not modify the forward-backward asymmetry.
In terms of the angular distribution this can be rephrased saying that
variations of the couplings $L_{9L},\: L_{9R}$ lead to a change of the overall
normalization of the differential cross-section, while changes in $\hat\alpha$
lead to changes in the shape of the distribution. This effect will be
demonstrated further when we discuss the angular distributions.

\subsection{$\chi^2$ Analysis of the Angular Distribution}

\noindent
In this section we discuss the bounds on the anomalous
couplings that can be obtained from the analysis of the differential
cross-section $d\sigma / d\cos\theta$. We will use the $\chi^2$
criterion in the form of Eq.~(\ref{chi2}) with experimental
uncertainties defined in Eq.~(\ref{uncer1}).
We will allow two parameters to vary at a time while fixing the third one
at its standard model value (0 at tree-level). Therefore, in order
to use a $\chi^2$ approach we  need a minimum of 4 bins
to have $N_{DOF} = N_{measurements}-N_{parameters}-1=1$.
We will consider the cases with
the angular region ($-0.989< \cos\theta < 0.4$) divided into 4, 5,
and 10 bins.
To compare these $\chi^2$ results with those obtained in  the previous
section using the criterion Eq.~(\ref{total2}),
we adopt the same C.L. of 95\%.

For the $\chi^2$ approach it is
important to understand which is the number of bins that gives the strongest
bounds on the parameters given an event sample.
As we mentioned before, the total expected
number of reconstructed $WW$-events for the chosen
luminosity is $\sim 5.5\times 10^4$. However,
with the kinematical cut on scattering angle that we use,
$-0.989 < \cos\theta < 0.4$, this number is reduced to
4384 events. With unpolarized beams and choosing 4 angular bins,
the number of events in each bin varies from 327 to 2175
(with the smaller number in the backward-most bin). These numbers
correspond to  relative  statistical errors varying from 3.8\%  to
2.1\%. For the case of 5(10) bins the number of events varies from
229(81) to 1854(1068), and the statistical error varies
from 6.6\%(11.1\%) to 2.3\%(3.1\%). If the beams are polarized
there is an even larger loss of statistics due to the partial cancellation
of the dominant $t$-channel diagram.
One can see that for these binnings of the events
the corresponding statistical errors  are larger than the systematic error.
This means that we have a statistically unsaturated event
sample, and the strongest bounds are obtained with the minimum number of bins.

Before using the angular distribution to place bounds on the parameters,
it is useful to see the behaviour of this distribution for small deviations
from the standard model. For illustration purposes we choose
the values $L_{9L}=5$, $L_{9R}=5$, and $\hat \alpha =5$. Notice that these
numbers are are small enough to
neglect the difference between the ``quadratic'' and ``linear''
approximations.

\begin{figure}[htb]
\centerline{\epsfxsize=2.in\epsfbox{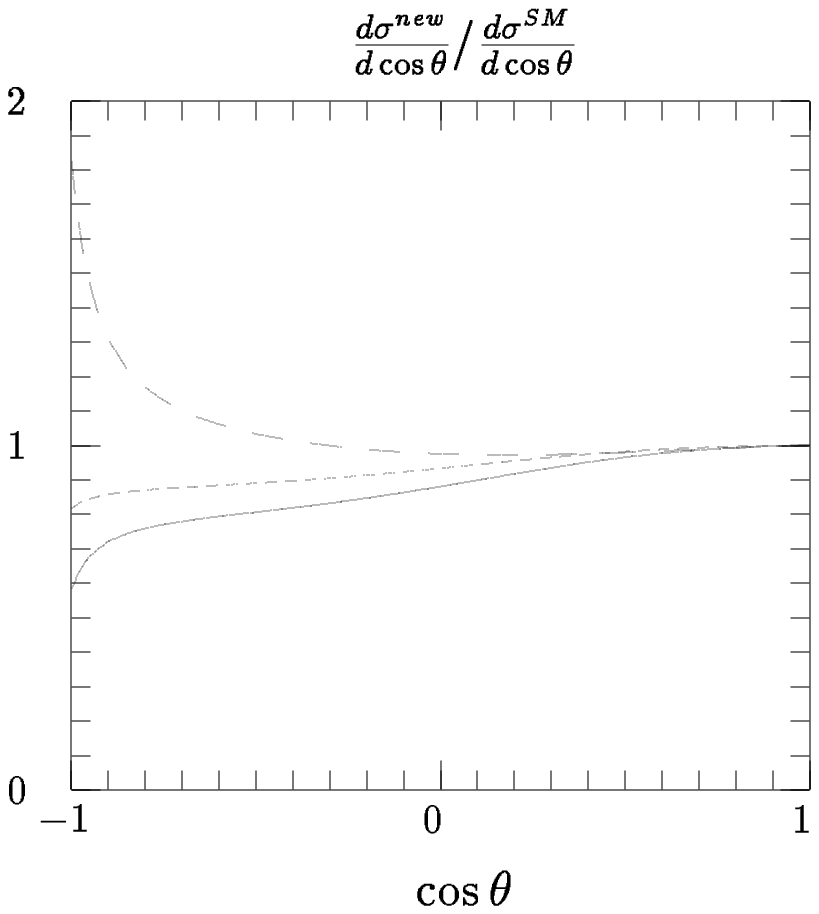}\hspace{0.5in}
\epsfxsize=2.in\epsfbox{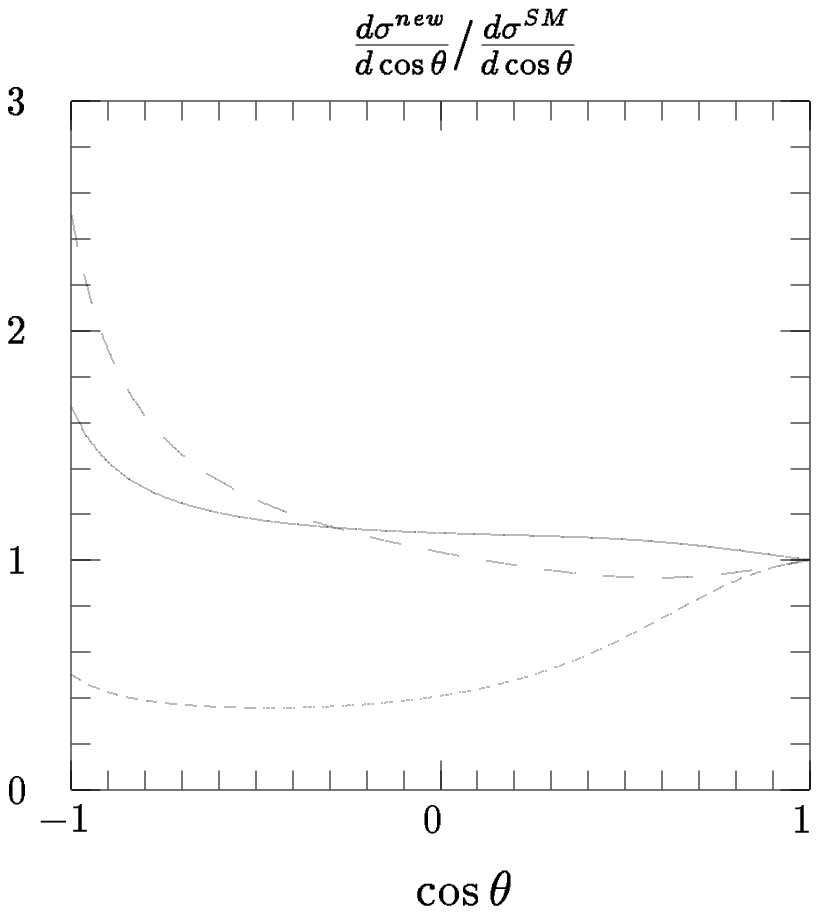}}
\caption{
Angular distributions normalized to the standard model for
a) unpolarized beams ($z_1=z_1=0.0$) and b) maximally polarized beams
($z_1=z_1=0.8$). The solid, short-dashed and long-dashed
lines correspond to $L_{9L}=5$, $L_{9R}=5$, and $\hat\alpha =5$
respectively.}
\label{fdisnor}
\end{figure}

In Fig.~\ref{fdisnor} we show the behaviour of the angular distribution for the
unpolarized case in the range $-0.989 < \cos\theta < 0.4$, normalized
to the angular distribution predicted by the standard model. The solid line
corresponds to $L_{9L}=5$, the short dashed line corresponds to $L_{9R}=5$,
and the long dashed line
corresponds to $\hat\alpha =5$.
In Fig.~\ref{fdisnor}a (\ref{fdisnor}b) we present the normalized
angular distributions for unpolarized (polarized) beams.
One can see in Fig.~\ref{fdisnor}a that
variations of $L_{9L}$ and $L_{9R}$ lead to a change in the overall
normalization of the distribution; whereas variations in $\hat\alpha$
result in a change in the shape of the distribution. However,
this difference is not evident in the case
of polarized beams (see Fig.~\ref{fdisnor}b).

\begin{figure}[htb]
\centerline{\epsfxsize=3.in\epsfbox{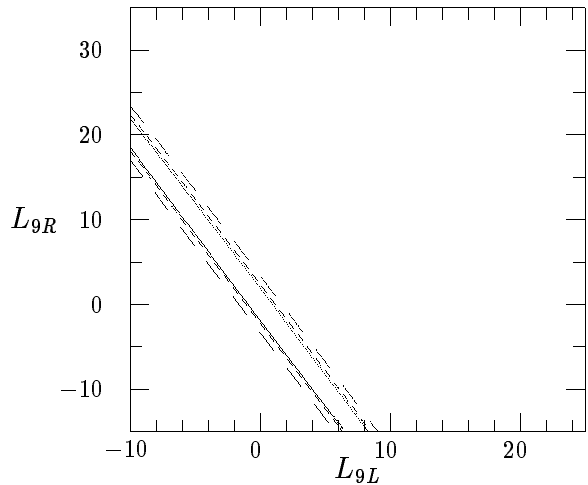}}
\caption{
$L_{9L}-L_{9R}$-projections of the allowed parameter region
(``linear'' approximation) for the
unpolarized case ($z_1=z_2=0.0$) corresponding to a 95\% C.L.
$\chi^2$-analysis for the cases of 4 (solid line), 5 (short-dashed line),
and 10 (long-dashed line) bins.}
\label{fcomp}
\end{figure}

In Fig.~\ref{fcomp} we show the projection of the allowed
parameter region in the $L_{9L}-L_{9R}$ plane for
unpolarized beams, which corresponds to 95\% C.L. in the $\chi^2$-analysis
for the cases of 4 (solid line), 5 (short-dashed line), and 10 (long-dashed
line) bins. One can see that the best bounds are, indeed, obtained with
the smallest number of bins, four. The same result holds true for
polarized beams.

\begin{figure}[htb]
\centerline{\epsfxsize=2.in\epsfbox{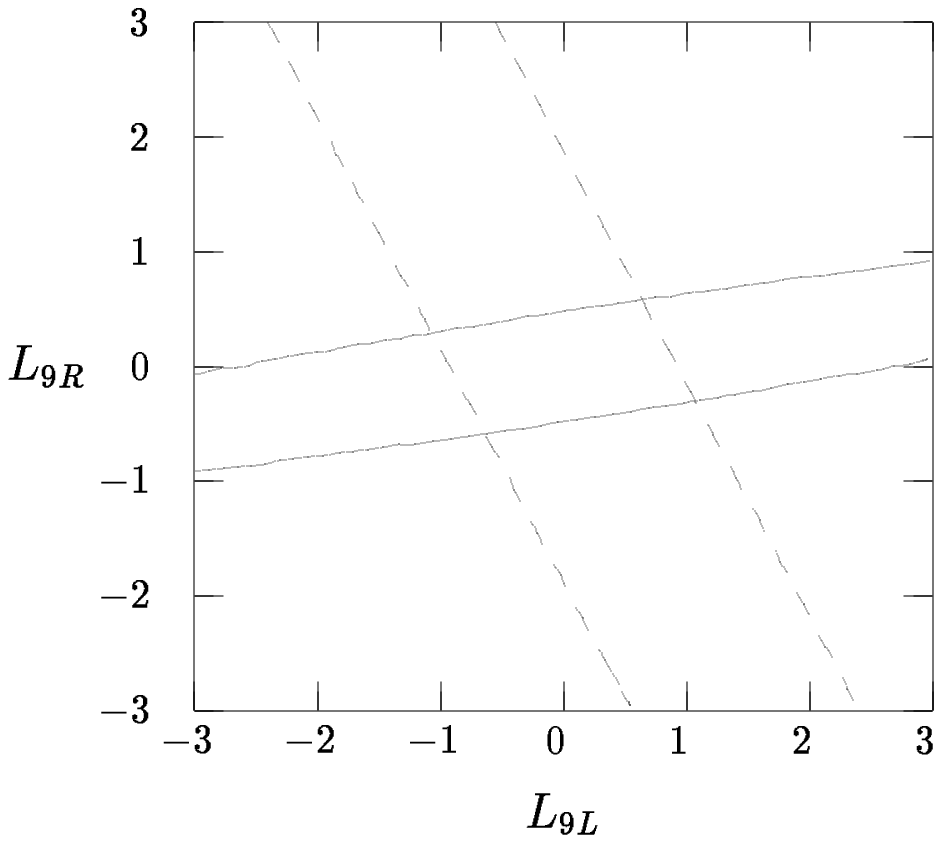}\hspace{0.5in}
\epsfxsize=2.in\epsfbox{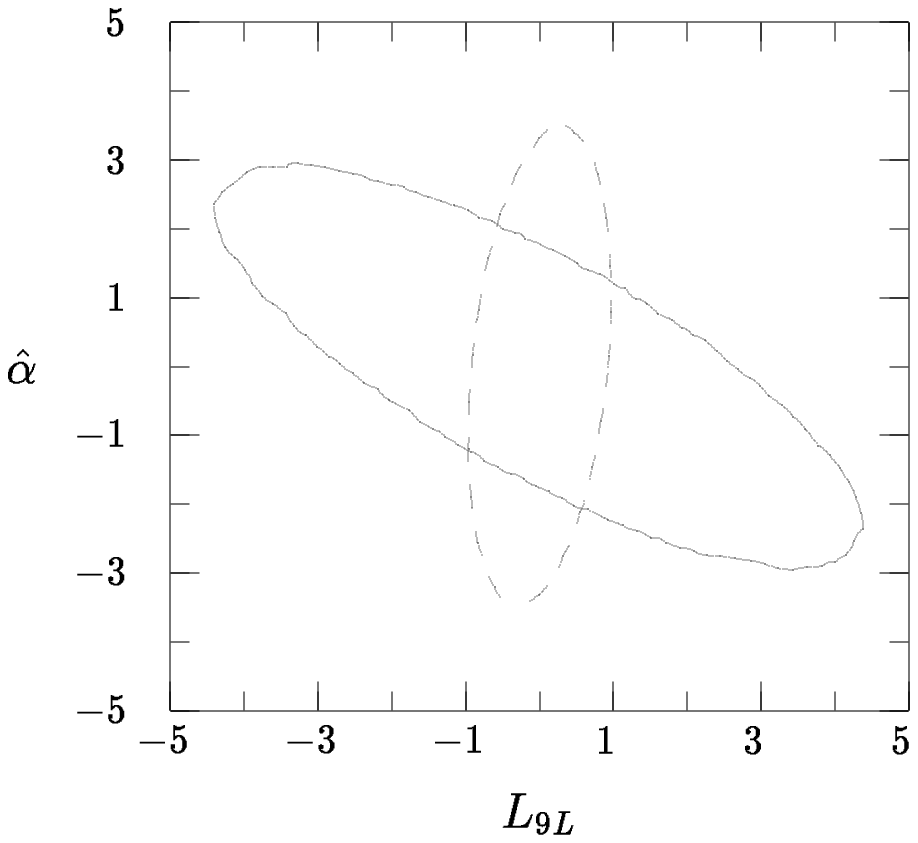}}
\centerline{\epsfxsize=2.in\epsfbox{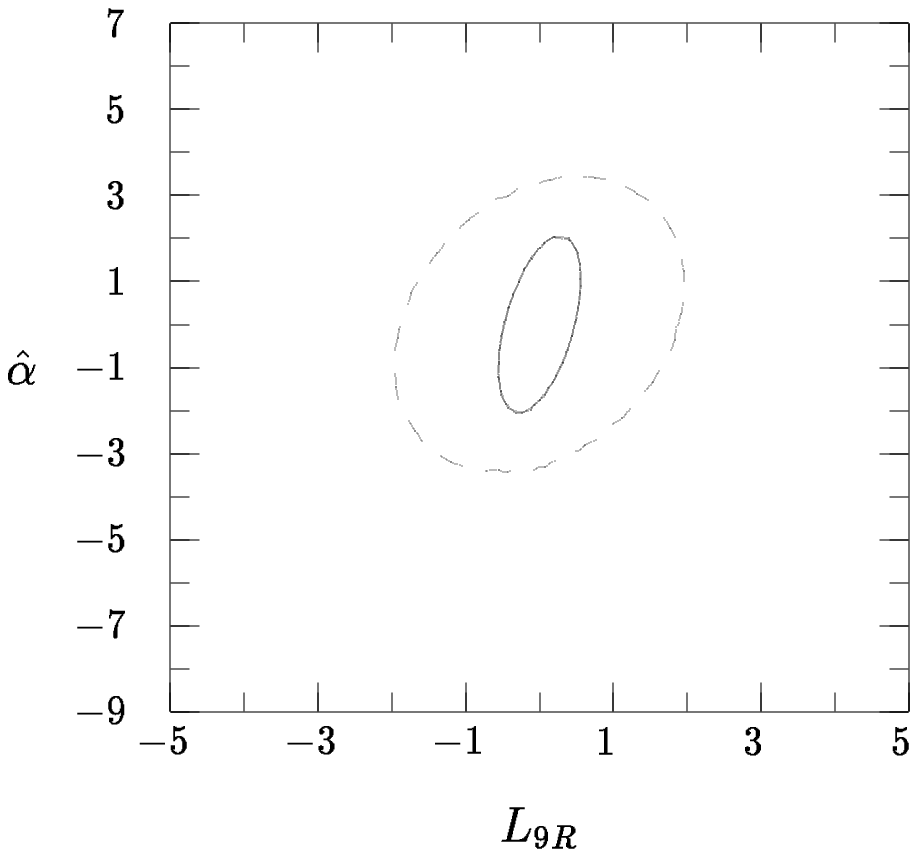}}
\caption{
Allowed regions (``linear'' approximation) from a $\chi^2$ analysis with
four bins. The dashed curves correspond to $z_1=z_2=0$ and the solid curves
to $z_1=z_2=0.8$.
   a)  $L_{9L}-L_{9R}$, when $\hat \alpha =0$;
   b)  $L_{9L}-\hat \alpha$, when $L_{9R}=0$;
   c)  $L_{9R}-\hat \alpha$, when $L_{9L}=0$.}
\label{fchi}
\end{figure}

We find that the angular distribution gives slightly better bounds than
the combined criterion of Eq.~(\ref{total2}), as shown in Fig.~\ref{fchi}.

Thus, choosing the case of 4 bins we can present the resulting bounds
on  $L_{9L}$, $L_{9R}$, and $\hat \alpha$  following from
the $\chi^2$-analysis of the angular distribution, which are shown in
Fig.~\ref{fchi}. The two-parameter fit bounds
(setting one of the three couplings at a time to its standard model value) are:
\begin{eqnarray}
-1.2\: \leq &L_{9L}& \leq \: 1.0\;, \nonumber\\
-0.6\: \leq &L_{9R}& \leq \: 0.7 \;, \label{bounds3} \\
-3.5 \: \leq &\hat \alpha & \leq \: 3.5\:. \nonumber
\end{eqnarray}

\section{Summary and Conclusions}

\noindent
If the electroweak symmetry breaking sector is strongly interacting, and
there are no new resonances below a TeV
one expects deviations of the gauge boson self-interactions from their
standard model values. In theories that conserve $CP$ and have an
approximate custodial symmetry we can parameterize these deviations in
terms of three constants, $L_{9L}$, $L_{9R}$ and $\hat\alpha$.
An $e^+e^-$ collider operating at $\sqrt{s}= 0.5$~TeV with
polarized beams and an integrated luminosity of 50~fb$^{-1}$
can provide important input into our understanding
of the nature of electroweak symmetry breaking.
We find that such a collider can place the following bounds:
\begin{eqnarray}
(-1.4 \to -1.2)\: \leq &L_{9L}& \leq \: (1.0\to 1.4)\;, \nonumber\\
(-0.7\to -0.6)\: \leq &L_{9R}& \leq \: 0.7 \;, \label{bounds5} \\
(-3.5\to -3.3)\: \leq &\hat \alpha & \leq \: (3.2\to 3.5)\:. \nonumber
\end{eqnarray}
The ranges correspond to the difference between the ``linear'' and
``quadratic'' approximations, and to the difference between using the
simple criterion of Eq.~(\ref{total2}) and a more sophisticated $\chi^2$
analysis of the angular distribution. These differences can be taken as
a rough guide of the theoretical uncertainties under our stated assumptions.

The authors of Ref.\cite{gouna} have also studied the process
$e^+e^- \ra W^+ W^-$ in terms of anomalous couplings at a future $e^+e^-$
collider like the one we discuss here. Because they do not have in mind
a strongly interacting electroweak  symmetry breaking sector, as we do,
they look for deviations of the standard model in terms of a larger number
of parameters than we do. They do not, however, study the parity
violating coupling $\hat{\alpha}$. A meaningful comparison of their
results with ours involves their two-parameter fit to their quantities
$\delta_Z$ and $X_\gamma$ which we translate into\footnote{Our
$\chi^2$ analysis is different from that of Ref.~\cite{gouna}, page 747.
Nevertheless, we take their results at face value to compare with our
results since their bounds would be weaker using our $\chi^2$ criterion
and our conclusion remains the same.}
\begin{eqnarray}
-2.0 \: \leq &L_{9L}& \leq \: 1.8\;, \nonumber\\
-3.4 \: \leq &L_{9R}& \leq \: 4.7 \;, \label{boundgou}
\end{eqnarray}
We can see that the bounds we obtained by combining unpolarized and
polarized collisions are significantly better. This is especially true
for the case of $L_{9R}$. This emphasizes the
additional sensitivity to new physics provided by polarized beams.

We have shown that polarized beams with adjustable
degrees of polarization would constitute a very significant tool
in the search for new physics. In terms of new physics parameterized
by a set of anomalous couplings, beam polarization makes it possible
to explore directions of parameter space that cannot be reached
in unpolarized collisions.

To place our bounds in perspective, we now compare them to those
obtained from LEP~I and those that can be obtained at LEP~II.
Precision measurements of $Z$ partial widths imply \cite{valen1}:
\begin{eqnarray}
-28\: \leq &L_{9L}& \leq \: 27\;, \nonumber\\
-9\: \leq &\hat{\alpha}& \leq \: 5\;, \nonumber\\
-100\: \leq &L_{9R}& \leq \: 190 \;. \label{lep1}
\end{eqnarray}
Expected bounds from LEP~II with $\sqrt{s}=190$~GeV and $\int {\cal L} dt =
500$~pb$^{-1}$ are \cite{boud}
\begin{eqnarray}
-41\: \leq &L_{9L}& \leq \: 26\;, \nonumber\\
-100\: \leq &L_{9R}& \leq \: 330 \;. \label{lep2}
\end{eqnarray}

Similar bounds have been obtained for different future colliders.
For example, with an $e\gamma$-collider with $\sqrt{s_{ee}}=
500$~GeV and $\int {\cal L} dt = 50$~fb$^{-1}$ they are \cite{valen2}:
\begin{eqnarray}
(-7 \to -5)\: \leq &L_{9L}& \leq \: (4 \to  6)\;, \nonumber\\
(-17\to -5)\: \leq &L_{9R}& \leq \: (4 \to 16) \;, \label{eg} \\
-15 \: \leq &\hat \alpha & \leq \: 7\:. \nonumber
\end{eqnarray}
Studies for the LHC (with $\sqrt s = 14$ TeV and integrated luminosity
100 fb$^{-1}$) have found \cite{group} a sensitivity to
$L_{9L}$ of order $10$.

After completion of this paper a similar analysis by M.~Ginter {\it et. al.}
has appeared \cite{newgodf}. These authors consider polarized
electron beams as we do, and they reach similar conclusions to ours
for the parameters that are common to our study\footnote{These are $L_{9L}$
and $L_{9R}$ albeit with a different normalization.} in the case of
one-parameter fits.

\section*{Acknowledgements}

\noindent
The work of A.~A.~L.
has been made possible by a fellowship of Royal Swedish Academy of Sciences
and is carried out under the research program of International
Center  for Fundamental Physics in Moscow. A.~A.~L.is also supported in part
by the International Science Foundation under grants NJQ000 and NJQ300.
The work of
T.H. is supported in part  by  the DOE grant  DE-FG03-91ER40674
and in part  by  a UC-Davis Faculty Research Grant. The work of G.V. was
supported in part by the DOE OJI program
under contract number DE-FG02-92ER40730.
We thank S. Dawson for useful discussions.

\appendix

\section*{Analytic expressions for the cross-section}

\noindent
We present below the explicit expressions for the dimensional functions \\
$T_{ij}=T_{ij}(M_W,\: \kappa_{\gamma,Z}, g_{1\gamma,1Z},\: g_5,\: s,\: t)$ used
in expressions
(\ref{sigmat}) for the cross-section of the $e^+e^-\to W^+W^-$
process. In this appendix we use $M\equiv M_W$, and $t$ is the absolute value
of the usual Mandelstam variable. Because we do not need to consider the
renormalization due to $L_{10}$ as explained in the text, the parameters
$a_f= T_{3f}/2c_\theta s_\theta$ and $v_f = (T_{3f}-2Q_fs^2_\theta)/
2s_\theta c_\theta$ are the usual tree-level standard-model axial and vector
couplings of the $Z$ to fermions.

\begin{eqnarray*}
 T_{11} = \frac{t^3}{3} &\cdot & (4 s M^2 g^2_{1\gamma}+4 s M^2
\kappa^2_{\gamma}
              -24 M^4 g^2_{1\gamma}-2 s^2 \kappa^2_{\gamma})\\
         -\frac{t^2}{2}&\cdot &(4 s^2 M^2 g^2_{1\gamma}+8 s^2 M^2
\kappa^2_{\gamma}
               -32 s M^4 g^2_{1\gamma}-8 s M^4 \kappa^2_{\gamma}
             +48 M^6 g^2_{1\gamma}-2 s^3 \kappa^2_{\gamma})\\
       +t &\cdot & (4 s^3 M^2 g_{1\gamma} \kappa_{\gamma}+2 s^3 M^2
g^2_{1\gamma}
           +2 s^3 M^2 \kappa^2_{\gamma}-16 s^2 M^4 g^2_{1\gamma}
\kappa_{\gamma}
           -8 s^2 M^4 g^2_{1\gamma}\\
           && -10 s^2 M^4 \kappa^2_{\gamma}
           +4 s M^6 g^2_{1\gamma}+4 s M^6 \kappa^2_{\gamma}-24 M^8
g^2_{1\gamma})
\end{eqnarray*}

\begin{eqnarray*}
T_{12}= \frac{t^3(v_e S_1-a_e S_2) }{3} &\cdot & (4 s M^2 g_{1Z}
g_{1\gamma}+4 s M^2
\kappa_Z \kappa_{\gamma}
             -24 M^4 g_{1Z} g_{1\gamma}-2 s^2 \kappa_Z \kappa_{\gamma})\\
      -\frac{t^2(v_e S_1-a_e S_2) }{2} &\cdot &(4 s^2 M^2 g_{1Z} g_{1\gamma}+
8s^2M^2 \kappa_Z\kappa_{\gamma}
               -32 s M^4 g_{1Z}g_{1\gamma}-8 s M^4 \kappa_Z\kappa_{\gamma}  \\
             &&  +48 M^6 g_{1Z} g_{1\gamma}-2s^3 \kappa_Z \kappa_{\gamma})\\
      +t(v_e S_1-a_e S_2) &\cdot & (2 s^3M^2 g_{1Z} g_{1\gamma}+2 s^3M^2
g_{1Z}\kappa_{\gamma}
             +2 s^3 M^2 \kappa_Zg_{1\gamma}+2s^3 M^2 \kappa_Z \kappa_{\gamma}\\
            && -8 s^2 M^4 g_{1Z} g_{1\gamma}-8s^2 M^4 g_{1Z}\kappa_{\gamma}
             -8 s^2 M^4 \kappa_Z g_{1\gamma}-10 s^2 M^4 \kappa_Z
\kappa_{\gamma}\\
             && +4 s M^6 g_{1Z} g_{1\gamma}+4 s M^6 \kappa_Z \kappa_{\gamma}
            -24 M^8 g_{1Z}g_{1\gamma})\\
      -\frac{t^2(a_e S_1-v_e S_2)g_5}{2} &\cdot & (4 s^2 M^2 g_{1\gamma}+4
s^2 M^2 \kappa_{\gamma}
            -16 s M^4 g_{1\gamma}-16 s M^4 \kappa_{\gamma})\\
     +t(a_e S_1-v_e S_2)g_5 &\cdot & (2s^3 M^2 g_{1\gamma}+2s^3 M^2
\kappa_{\gamma}
           -12s^2 M^4 g_{1\gamma}-12s^2 M^4 \kappa_{\gamma}
           +16 s M^6 g_{1\gamma}\\
            &&+16 s M^6 \kappa_{\gamma})
\end{eqnarray*}

\begin{eqnarray*}
 T_{13} = \frac{t^3}{3}&\cdot & (4 M^2 g_{1\gamma}-2 s \kappa_{\gamma}) \\
        - \frac{t^2}{2} &\cdot &(4 s M^2 g_{1\gamma}+4 s M^2 \kappa_{\gamma}
             -2 s^2 \kappa_{\gamma})\\
        + t &\cdot & (4 s^2 M^2 g_{1\gamma}+4 s^2 M^2 \kappa_{\gamma}
              -10 s M^4 \kappa_{\gamma}-12 M^6 g_{1\gamma})\\
        - \ln \biggl({t \over 1{\rm ~GeV}^2}\biggr)
           &\cdot &(8 s M^6 g_{1\gamma} + 8 s M^6 \kappa_{\gamma}
             +8 M^8 g_{1\gamma})
\end{eqnarray*}

\begin{eqnarray*}
T_{22} = \frac{t^3( (v^2_e+a_e^2)S_1-2v_e a_e S_2)}{3} &\cdot & (4 s
 M^2 g^2_{1Z}
             +4 s M^2 \kappa^2_Z -24 M^4 g^2_{1Z}-2 s^2 \kappa^2_Z)\\
  -\frac{t^2( (v^2_e+a_e^2)S_1-2v_e a_e S_2)}{2} &\cdot &
(4 s^2 M^2 g^2_{1Z}
               +8 s^2 M^2 \kappa^2_Z
              -32 s M^4 g^2_{1Z}-8 s M^4 \kappa^2_Z \\
             && +48 M^6 g^2_{1Z}-2 s^3 \kappa^2_Z)\\
    +t( (v^2_e+a_e^2)S_1-2v_e a_e S_2) &\cdot & (4 s^3
 M^2 g_{1Z} \kappa_Z+2 s^3 M^2 g^2_{1Z}
             +2 s^3 M^2 \kappa^2_Z \\ && -16 s^2 M^4 g_{1Z} \kappa_Z
             -8 s^2 M^4 g^2_{1Z}-10 s^2 M^4 \kappa^2_Z
             +4 s M^6 g^2_{1Z}\\
             &&  +4 s M^6 \kappa^2_Z-24 M^8 g^2_{1Z}) \\
      -\frac{t^2g_5(2v_e a_e S_1-(v^2_e+a_e^2)S_2)}{2} &\cdot &
(8 s^2 M^2 g_{1Z}  +8 s^2 M^2 \kappa_Z -32 s M^4 g_{1Z}-32 s M^4 \kappa_Z) \\
      +tg_5 (2v_e a_e S_1-(v^2_e+a_e^2)S_2)&\cdot & (4 s^3 M^2 g_{1Z}
            +4 s^3 M^2 \kappa_Z-24 s^2 M^4 g_{1Z}
            -24 s^2 M^4 \kappa_Z\\  &&+32 s M^6 g_{1Z}+32 s M^6 \kappa_Z)\\
 + \frac{t^3g^2_5( (v^2_e+a_e^2)S_1-2v_e a_e S_2)}{3} &\cdot &
(4s M^2  -16 M^4)\\
 - \frac{t^2g^2_5( (v^2_e+a_e^2)S_1-2v_e a_e S_2)}{2}
&\cdot &(4s^2M^2
  -24s M^4 +32 M^6)\\
 +tg^2_5( (v^2_e+a_e^2)S_1-2v_e a_e S_2) &\cdot & (2 s^3 M^2
-16 s^2 M^4+36s M^6-16M^8)
\end{eqnarray*}

\begin{eqnarray*}
   T_{23} =  \frac{t^3}{3} &\cdot & (4 M^2 g_{1Z}-2 s \kappa_Z) \\
          - \frac{t^2}{2} &\cdot & (4 s M^2 \kappa_Z +4 s M^2 g_{1Z}-2
s^2 \kappa_Z) \\
          + t &\cdot &(4 s^2 M^2 \kappa_Z + 4 s^2 M^2 g_{1Z}-10 s M^4 \kappa_Z
-12 M^6 g_{1Z})\\
          - \ln  \biggl({t \over 1{\rm ~GeV}^2}\biggr)
           &\cdot & (8 s M^6 \kappa_Z+8 s M^6 g_{1Z}+8 M^8 g_{1Z})\\
          - \frac{t^2 g_5}{2} &\cdot & (8 s M^2-8 M^4)\\
          + t g_5 &\cdot &(4 s^2 M^2-8 s M^4+16 M^6) \\
          - \ln \biggl({t \over 1{\rm ~GeV}^2}\biggr)
           g_5 &\cdot &(8 s M^6-8 M^8)
\end{eqnarray*}

\begin{displaymath}
      T_{33} =-\frac{2}{3} t^3-\frac{1}{2} t^2 (4 M^2-2s)
          +t(8 s M^2-10 M^4)-\ln \biggl({t \over 1{\rm ~GeV}^2}\biggr)
           (-8 s M^4+16 M^6)
          +\frac{8}{t} M^8
\end{displaymath}

\end{document}